\documentclass[superscriptaddress,aps,pra,twocolumn,amssymb,amsfonts,letterpaper]{revtex4}
\usepackage{amsmath} \usepackage{amsthm} \usepackage{color}
\usepackage{graphics,graphicx}
\usepackage[pdftex]{hyperref}
\usepackage{epsfig}

\input{Qcircuit}

\def\cO{{\cal O}}

\def\({\left(}
\def\){\right)}
 
\newcommand{\braket}[2]{\langle#1|#2\rangle}

\newcommand{\sket}[1]{| #1\rangle}
\newcommand{\sbra}[1]{\langle #1|}

\begin{document}

\title{Necessary Condition for the Quantum Adiabatic Approximation}

\author{S. Boixo} \affiliation{Institute for Quantum Information,
  California Institute of Technology, Pasadena, CA 91125, USA}
\email{boixo@caltech.edu}

\author{R. D.  Somma}
\affiliation{Los Alamos National Laboratory, Los Alamos, New Mexico 87545, USA}
\email{somma@lanl.gov}

\date{\today}
\begin{abstract}  
A gapped quantum system that is adiabatically perturbed
remains approximately in its eigenstate after the evolution.  
We prove that, for constant gap, general quantum
processes that approximately prepare the final eigenstate 
require a minimum time proportional to the ratio of the length
of the eigenstate path to the gap. 
Thus, no rigorous adiabatic condition can yield a smaller cost. We also give a necessary
condition for the  adiabatic approximation that depends on local properties of the path, which is appropriate when the gap varies.

\end{abstract}

\maketitle
The quantum adiabatic theorem
asserts that a continuously perturbed and gapped 
quantum system remains in its instantaneous
eigenstate in the limit where the rate of change
of the perturbation vanishes~\cite{messiah_quantum_1999}. 
This assertion is {\em quantified}  via the adiabatic 
approximation~\cite{born_adiabatic_1928}, which provides a relation 
between the rate of change of the perturbation and the fidelity of 
the evolved state with the 
final eigenstate.  The adiabatic approximation 
is a key part of 
quantum computing as it determines the complexity of several
quantum algorithms~\cite{farhi_quantum_2000,combined1,somma_quantum_2008,wocjan_quantum_2008}.
In fact, adiabatic quantum computation~\cite{farhi_quantum_2000}, in which the result of a 
problem is encoded in the ground state of a (final) Hamiltonian, is equivalent to standard quantum 
computation~\cite{aharonov_adiabatic_2007}. Further,
adiabatic approximations play an important role in  
areas like Born-Oppenheimer theory, 
the quantum Hall effect and STIRAP~\cite{teufel_adiabatic_2003}.

 It is important to remark that some familiar adiabatic approximations 
 are known to be insufficient~\cite{combined2}
and sometimes unnecessary~\cite{combined3,boixo_quantum_2009}.  
The growing interest on the adiabatic approximation 
has spurred work on corresponding
rigorous conditions~\cite{combined3,combined3a}. 
In this manuscript we give a rigorous lower bound for the 
evolution time (or cost) of adiabatic processes that prepare the final state.
This bound is also valid for more general quantum evolutions~\cite{somma_quantum_2008,wocjan_quantum_2008,boixo_quantum_2009,boixo_optimal_2009}.

Let $\{H(r)\}$, with $r \in [0,1]$, be a given continuous Hamiltonian path 
and $\{\ket{\psi(r)}\}$ the corresponding non-degenerate
eigenstate path (eigenpath).
Adiabatic evolutions aim to prepare $\ket{\psi(1)}$ at bounded precision
from $\ket{\psi(0)}$ by choosing a proper schedule $r(t)$.
We recently argued~\cite{boixo_quantum_2009} that the relevant quantities
for the adiabatic approximation are not only the minimum eigenvalue gap
of the Hamiltonians, $\Delta$, but also the length of the path to be traversed,
$L$. We presented a method
that adiabatically prepares the final state by evolving with the Hamiltonians
for suitable random times~\cite{boixo_quantum_2009}. The average cost of the randomization method is
$\cO (L^2/\Delta)$ when
the rate of change of the eigenstates along the path and the corresponding 
eigenvalues are known. 
A more efficient method to traverse the eigenpath for this case was introduced in~\cite{wocjan_quantum_2008}. This method uses
Grover's fixed point search and the schedule $r(t)$ is non-monotonic. It results in a
cost $\tilde \cO(L (\log L)^2/\Delta)$. (We use the soft order
notation $\tilde \cO$ to hide doubly logarithmic factors.)
Finally, we recently
derived a non-monotonic quantum algorithm or process
that dynamically estimates the rate of change
of the eigenstates, and results in a cost $\cO( L \log L/\Delta)$ under broader
assumptions~\cite{boixo_optimal_2009}.

Here we also consider  {\em general} quantum processes
that prepare the final eigenstate from the initial one,
at bounded precision, by evolving with the Hamiltonians.  
Thus, we do not exploit the unknown structure
of   $\{H(r)\}$, rather we work in the so-called
black-box model where the only assumption is to be able
to evolve with $H(r(t))$ for some schedule $r(t)$.
We then prove 
that the cost of such processes is, at least,  $\cO(L/\Delta)$.

To prove such a lower bound on the cost we introduce
particular instances of Hamiltonian paths $\{ H(r) \}$, and reduce them
to problems for which a query-complexity bound is known or
can be easily obtained. For example, to show the scaling with
$\Delta$, we can simply consider the adiabatic version
of Grover's search~\cite{grover_fast_1996,farhi_quantum_2000,roland_quantum_2002,boixo_quantum_2009}.
If $N$ is the problem size, the minimum gap is $1/\sqrt{N}$ and
$L \le \pi$ in this case. 
The lower bound $\sqrt{N} \in \cO(1/\Delta)$ for the search problem
is a celebrated result in
quantum computation~\cite{bennett_strengths_1997}. 
Showing the dependence of the cost on the path length and the minimum gap
requires a different analysis that constitutes our main contribution. 

For the instances considered below, the relevant Hamiltonian-eigenvalues,
gaps, and rates of change of the eigenstates are known and remain constant
for all $r$. %Most of these quantities are usually considered unknown in the derivation of adiabatic approximations.  
We also clarify
that a better bound may be obtained if 
these quantities vary along 
the path.
Still, we show that any rigorous adiabatic
approximation based on local properties of the path cannot
yield a schedule that satisfies 
 $  \dot r(t) > c \Delta(r)/\| \! \ket {\partial_r \psi(r)} \! \| $
for all $r$, and specific $c > 0$ given below.

Before obtaining the necessary condition for the adiabatic
approximation we give a precise definition of $L$  and comment on
the resulting cost for the 
worst-case. In the instances considered below the path length is  
\begin{align}
\label{eq:pl} 
L = \int_0^1
  \|\,\ket{\partial_r \psi(r)}\,\| d r\;.
\end{align}
(See~\cite{boixo_quantum_2009} for a general definition of $L$.)
With no loss of generality we assume
$\braket{\partial_r\psi(r)}{\psi(r)} =0$.
$L$ is the only natural length in projective
Hilbert space (up to irrelevant normalization factors). 
An upper bound on  $L$ is   $\| \dot H \|/\Delta$, with $\| \dot H \| = \max
\| \partial_r H \|$, and $\| . \|$ the operator norm. In the worst-case this 
bound is tight
and $L/\Delta \sim  \| \partial_r H \|/\Delta^2$. 
However, in many cases of interest $L$ can be bounded
independently of $\Delta$ and the algorithms
in~\cite{boixo_quantum_2009,wocjan_quantum_2008,boixo_optimal_2009}
result in much smaller implementation costs than those determined by other
rigorous adiabatic
approximations~\cite{combined3,combined3a}. 

The remainder of this manuscript is organized as follows.
We first give a simple proof of the necessary condition
for the adiabatic approximation when the eigenstates
are degenerate. This proof will motivate the desired
result  in the non-degenerate case, proven later.

\begin{figure}[ht]
  \centering
  \includegraphics[width=4cm]{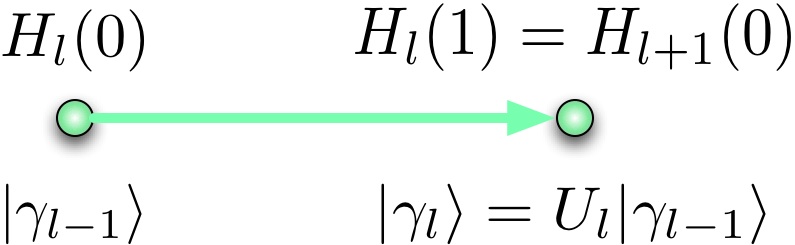}
 \caption{A quantum process that traverses the $l-$th path segment 
  applies $U_l$.}
  \label{fig:segment}
\end{figure}

{\bf Eigenpath within a degenerate subspace.} 
We use constant-gap piece-wise adiabatic universal computation~\cite{combined4}. 
Let  $U=U_{n} \cdots  U_1$ be a unitary quantum circuit acting on a Hilbert space
$\cal H$, initially in $\ket \psi$. For $l \ge 1$, we define
the parametrized Hamiltonians (see Fig.~\ref{fig:segment})
\begin{align}
 H_l (s)&= - \frac \Delta 2  \left[ \cos(\pi s)( \openone \otimes (\ket {l-1} \bra {l-1}   - \ket{l} \bra {l}) \right. \\ & 
  \quad + \sin(\pi s) (U_l \otimes
 \ket {l} \bra{l-1} +  U_l^\dagger \otimes \ket {l-1} \bra {l}) ]  \nonumber \;, 
\end{align}
acting on ${\cal H} \otimes {\cal I}$, with  $s \in [0,1]$ and $\Delta$ the relevant gap.
The Hilbert space $\cal I$
encodes the step of the circuit using the so-called {\em clock} 
states~\cite{aharonov_adiabatic_2007,combined5}. 
For arbitrary $\ket \psi$, we define $ \ket{\gamma_l} = (U_l \cdots U_1 \ket {\psi})\ket{l}$
and $\ket{\gamma_0}= \ket \psi \ket 0$. 
Each $H_l(s)$ leaves invariant the subspace spanned by $\{\ket{\gamma_{l-1}},\ket{\gamma_l}\}$.
The eigenstates of $H_l(s)$ are degenerate; however, we are interested in the 
continuous path determined by the eigenstates
$\sket{\phi^l(s)} = \cos(\pi s/2) \ket {\gamma_{l-1}}
+ \sin (\pi s/2) \ket {\gamma_l}$.

The continuous Hamiltonian path $\{H(r)\}_{0 \le r \le 1}$ is constructed
by concatenation of $n$ path segments, each segment linking $\ket{\gamma_{l-1}}$ to $\ket{\gamma_{l}}$. 
Then $\ket{\psi(1)} = (U_n \cdots U_1 \ket{\psi}) \ket{n}$ and
the final eigenstate of the path contains
the state prepared by the circuit. 
The $l-$th path segment, where  $(l-1)/n \le r < l/n$,
 is determined by $H(r)=H_l(rn-l+1)$
(Fig.~\ref{fig:segment}).  
The relevant eigenstate for this path is
 $\ket{\psi(r)}= \sket{\phi^l(rn-l+1)}$.  
 Because the rate of change of 
the eigenstate is constant in the intervals $(l-1)/n \le r < l/n$,
and each path segment occurs in a two-dimensional subspace,  the
path length is
$L = \pi n$.  The adiabatic procedure considered is universal for 
quantum computation.

No general unitary
quantum process that evolves with the Hamiltonians
$\{H(r)\}$ for some schedule $r(t)$, and interleaves these evolutions with other (known)
operations, can prepare the final state at bounded precision with cost less
than $\cO(L/\Delta) \in \cO(n/\Delta)$. This lower bound easily follows from 
considering those quantum circuits $U$ built from particular instances of 
Grover's search. Each operation $U_j$ that composes the circuit is then 
a so-called Grover iteration~\cite{grover_fast_1996} which uses one {\em search} query.
The size of the circuit is $n=\sqrt{N}$, where $N$ is the size of the problem. 
In~\cite{farhi_analog_1996} it was shown that the cost of any 
continuous-time quantum algorithm that uses the Hamiltonians $H(r)$
and outputs the desired state
is, at least, $T \in \cO(\sqrt{N}/\Delta)$;
otherwise the search problem could be solved using less than $\sqrt{N}$ queries.
Since
$L \in \cO(\sqrt{N})$, this proves 
the result. A basis for the eigenspace of the final Hamiltonian is 
$\{\ket{\chi}\ket{n}\}_{\chi}$, and thus reaching the
final eigenspace is easier than drawing up the right state.

{\bf Non-degenerate eigenpath.} We now consider the case where the eigenstates are non-degenerate.  We show that, for any given $\{H(r)\}$ and $\Delta$, no generic evolution induced by a Hamiltonian $H(r(t)) +H_D(t)$, where $H_D(t)$ is a driving Hamiltonian, can prepare an approximation to the final eigenstate in time less than $\cO(L/\Delta)$.  The result is valid for any $r(t)$ and $H_D(t)$ that do not depend on the unknown structure of $H(r)$, but may depend on the known gap and rate of change of the eigenstates~\cite{comment1}. This information is usually not available when deriving adiabatic approximations; our lower bound on the cost clearly encompasses those cases as well.  Our setup is general, as it may include controlled-Hamiltonian evolutions as well as intermediate measurements.  To prove the result we consider instances that reduce to instances of the ordered search problem~\cite{combined6}.  A discrete-query complexity lower bound, based on the {\em adversary method}~\cite{bennett_strengths_1997,hoyer_quantum_2002,combined7}, is known for this problem. We extend this result to the continuous-time query setting.

Let $x=[x_0, \ldots ,x_{n-1}]$, $x_j \in \{0,1\}$, be a {\em secret word} (input)
and define the Hamiltonians
\begin{align} \label{eq:oo}
H_{x}^l =   \Delta \sket{ x(l),+,\ldots,+} \sbra{x(l),+,\ldots,+} \; ,
\end{align} 
acting on $n$ qubits, where $x(l)=[x_0,\ldots,x_{l-1}]$,
and $\ket{+}=(\ket{0}+\ket{1})/\sqrt{2}$.
The evolution induced by  $H_{x}^l$ for time $\pi/\Delta$
is equivalent to a phase query $Q_x^l$ that, for input state $\ket{a_0, \ldots, a_{l-1}}$,
outputs $- \ket{a_0, \ldots, a_{l-1}}$ when $a_j = x_j \ \forall \ 0 \le j \le l-1$, or 
does nothing otherwise. Basically, with one query $Q_x^l$ we answer the question:
{\em are the first $l$ bits of $x$ equal to $[a_0,\ldots,a_{l-1}]$}?.
The Hamiltonian path we will consider is one that has $\sket{x(l),+,\ldots,+}$
as intermediate eigenstates. Then $L \in \cO(n)$.
As in the degenerate case,
this path is built upon $n$ segments, each interpolating $H_x^l$
with $H_x^{l+1}$ (see below). A measurement
on the final state allows us to learn the input.

We now reduce our problem to the ordered search problem. 
The goal of ordered search is to find 
the input $x$ that represents a {\em marked item}
in an ordered list of $N=2^n$ elements.   A query  in this case,
acting on input $a=[a_0,\ldots,a_{n-1}]$, indicates whether 
the marked element $x$ is before position $a$ or not. 
The corresponding phase query $R_x$ puts a $\pm$
sign on the input state depending on the answer (see below).
It requires  $\cO(n)$ queries of type $R_x$ to solve the 
problem~\cite{hoyer_quantum_2002}. We remark that a query of
type $Q_x^l$ can be implemented using two queries of type $R_x$:
to decide if the first $l$ bits of the secret word coincide with $a(l)=
[a_0,\ldots,a_{l-1}]$, 
it suffices to call $R_x$ with inputs $[a_0,\ldots,a_{l-1},0,\ldots,0]-1$ 
(binary subtraction) and
$[a_0,\ldots,a_{l-1},1,\ldots,1]$, respectively. If the corresponding $R_x$'s outputs 
are different, then $a(l)=x(l)$. Then,
\begin{align}\label{eq:qx}
  Q_x^l = U^l R_x V^l R_x W^l\;, 
\end{align}
where $U^l$, $V^l$, 
and $W^l$ are the $x$-independent 
unitaries used to build a circuit that simulates
$Q_x^l$ using $R_x$. 

A simple argument based on the above analysis 
intuitively explains our result. It is now clear that no 
general quantum process that uses type-$Q_x^l$ queries can 
find the secret word with less than $\cO(n)$ queries;
otherwise, by replacing the queries, we would solve ordered search with less
than $\cO(n)$ type-$R_x$ queries. Since
each type-$Q_x^l$ query can be implemented
using the Hamiltonians $H_x^l$ for time $\pi/\Delta$,
 it is plausible that the total cost
of a continuous-time quantum process that
uses $H_x^j$ is at least $\cO(n/\Delta) \in \cO(L/\Delta)$. 
In fact, using the equivalence between continuous- and discrete-time query models
in~\cite{cleve_efficient_2008} yields a
lower bound for the cost $\tilde \cO[L/(\Delta \log (L/\Delta))]$. 
Our formal proof below, that uses 
a version of the adversary method
in the continuous-time
setting, will avoid the logarithmic correction in the cost.

Let 
$ H_x^l (s)=\cos(\pi s/2) H_x^{l-1}+\sin(\pi s/2) H_x^l \; .$
We build a particular Hamiltonian path as  in the degenerate case:
for $(l-1)/n \le r \le l/n$ we set $H_x(r)=H_x^l(rn-l+1)$. 
The eigenstates $\ket{\psi(r)}$ of $H_x(r)$,
with lowest eigenvalue, are non-degenerate. Further, the gap 
of $H_x(r)$ is $\Delta$ for all $r$.
A quantum algorithm that traverses the eigenpath will aim to prepare 
$\sket{x(l),+,\ldots,+}$ from $\sket{x(l-1),+,\ldots,+}$, learning a bit
of information about the secret word at each path segment.  
The path length is 
$L = \pi n/2$. 

We consider evolutions with Hamiltonian $H_x(r(t)) + H_D(t)$, where  $r(t)$ 
and $H_D(t)$ do not depend on $x$, the only unknown
quantity in this case. Ideally, after some
time $T>0$, the initial state approximately evolves to 
the desired eigenstate $\ket{\psi(1)}=\ket{x_0,\ldots,x_{n-1}}$.
Evolutions of this type include
continuous-time processes based on eigenpath
traversal, such as adiabatic evolutions where $H_D=0$.
If $\ket{\phi_x(t)}$
is the evolved state,
\begin{align} 
\label{eq:schro}
  i \,\partial_t \ket{\phi_x(t)} = (H_x(r(t)) + H_D(t)  )\ket{\phi_x(t)}\;.
\end{align} 
The initial state
$\ket{\phi_x(0)}$ is also independent of $x$.

We use the adversary method to
show that $T$ is at least $\cO(n/\Delta)$. 
At its core, the adversary method 
provides a bound for the rate of change of the overlap between
evolutions corresponding to different inputs. 
To distinguish between these inputs,
the evolved states must satisfy
$|\braket{\phi_x(T)}{\phi_y(T)}| \le \epsilon$,
for some $T>0$ and small $\epsilon$.

Let $\Gamma$ be an
{\em adversary matrix}: an irreducible symmetric matrix with
non-negative entries and zeros in the
diagonal. Denote by $\|\Gamma\|$ its operator norm, and by $v$ the principal unit
eigenvector, $\Gamma v = \| \Gamma \| v$. The following function serves as a measure of the {\em distinguishability} 
between evolutions with different inputs:
\begin{align}\label{eq:wt}
  W(t) = \sum_{x,y} \Gamma_{x,y} v_x v_y \braket{\phi_x(t)}{\phi_y(t)}\;.
\end{align}
 Since the initial state is independent of the input,
$W(0)=\| \Gamma \|$. Moreover, $W(T) \le \epsilon \| \Gamma \|$.

An upper bound on $| \partial_t W(t) |$ can be obtained
if we extend known results for the ordered search problem
to our case. We write the Hamiltonians $H_x^l$ as (see Eq.~\eqref{eq:qx})
\begin{align}
H_x^l =   \Delta (\openone -  Q_x^l) =   \Delta ( \openone - U^l R_x V^l R_x W^l) \; .
\end{align}
Then, using Eq.~\eqref{eq:schro} and the triangle inequality,
\begin{align}
\label{eq:b1}                   %
 | \partial_t \braket {\phi_x(t)}{\phi_y(t)} |  &\le  \max_l |\!
  \bra{\phi_x(t)} H_x^l \!-\! H_y^l \ket{\phi_y(t)}\! | \\
  \nonumber 
& \le  2 \Delta \max_l |\!
  \bra{\phi_x(t)} U^l R_x V^l R_x W^l - \\
  \nonumber  & \quad \quad \quad \quad -U^l R_y V^l R_y W^l \ket{\phi_y(t)}\! | \; .
\end{align}

For each $x$, we define a $2^n$-bit string $\alpha_x$ with entries 
$\alpha_x^i=0$, if $i<x$, and $\alpha_x^i=1$ otherwise. Let $P_i = \ket i
\bra i$ and  write
$\sbra{\phi_x^l(t)}= \sbra{\phi_x(t)} U^l R_x V^l $, and 
$\sket{\tilde \phi_y^l(t)}= V^l R_y W^l \ket{\phi_y(t)}$. The triangle inequality
and Eq.~\eqref{eq:b1} yield
\begin{align}
\nonumber
 | \partial_t \braket {\phi_x(t)}{\phi_y(t)} |    
& \le  2 \Delta \max_l \sum_{i: \alpha_x^i \ne \alpha_y^i}[ |\!
  \sbra{\phi_x^l(t)}  P_i W^l \ket{\phi_y(t)} \! |  \\
  \nonumber  & \quad \quad  \quad  \quad +| \! \bra{\phi_x(t)} U^l P_i  \sket{\tilde \phi^l_y(t)}\! | ] \; .
\end{align}
For $i \in \{0,\ldots,2^n-1\}$, we  introduce the matrices $\Gamma^i$ as
\begin{align}
  \Gamma^{i}_{x,y} = \left\{
      \begin{array}{ll}
        \Gamma_{x,y} & \textrm{if $\alpha_x^i \ne \alpha_y^i \; ,$}    \\
        0 & \textrm{otherwise \; .}
      \end{array}\right.
\end{align}
Then,
\begin{align}
  \nonumber 
| \partial_t W(t) |  &\le 2 \Delta \sum_{i,x,y} \Gamma_{x,y}^i v_x v_y
 \max_l [ |\!
  \bra{\phi^l_x(t)} P_i W^l \ket{\phi_y(t)}\! | \\
\nonumber 
& \quad \quad \quad \quad \quad + |\! \bra{\phi_x(t)} U^l P_i  \sket{\tilde \phi^l_y(t)}\! | ]  \\
\label{eq:b3}
& \le 4 \Delta \max_i \| \Gamma^i \| \; .
\end{align}
In addition, $ (1-\epsilon) \| \Gamma \|  \le W(0) - W(T)$ and
\begin{align}
\label{eq:inter}
 W(0) - W(T)
  & \le \int_0^T |\partial_t W(t)| dt \\
  \nonumber
 & \le  T 4 \Delta \max_{i} \| \Gamma^{i} \| \; .
\end{align}

The spectral lower bound for the cost
of the process that approximates the final eigenstate is $T\in \cO\( \| \Gamma\| /(4 \Delta  \max_{i} \| \Gamma^{i} \| ) \)$. 
We use the  adversary matrix for the ordered search problem~\cite{hoyer_quantum_2002}:
\begin{align}
\label{eq:advmatrix}
\Gamma_{x,y}   = \left\{
  \begin{array}{ll}
        \frac 1 { {\rm Hd}(\alpha_{x},\alpha_{y})} & \textrm{if $\alpha_{x} \ne \alpha_{y}$ ,} \\
        0 & \textrm{otherwise}
      \end{array}\right.
\end{align}
where ${\rm Hd}(\alpha_x,\alpha_y)$ is the 
Hamming distance. This choice 
yields $\| \Gamma \| \ge n$ and  $\max_{i} \| \Gamma^i \| \le \pi$.
Thus, 
$T  \in \Omega \( n/(4 \pi \Delta)  \)$. Since $L =\pi n /2$,
this proves the result
$T \in \cO(L/\Delta)$. 

With a similar construction  
we can prove a lower bound on the cost of
continuous-time query algorithms that
solve the ordered search problem, obtaining
$T \in \Omega(n/(2 \pi))$ in this case.

In the derivation of our result we used the fact that
the gaps of $\{H_x(r)\}$ are constant along the path. 
Nevertheless, the Hamiltonians in the path may have different (known) gaps
and one could be interested in
designing algorithms for eigenpath
traversal with a schedule $r(t)$ that depends
on the {\em local} gap $\Delta(r)$.
Consider again the instances above and 
redefine $H_x(r) \gets q(r) H_x(r)$, for known $q(r)>0$.
The new Hamiltonians have gaps $\Delta(r)= q(r) \Delta$.
We can replace $\Delta$ by $\Delta(r(t))$ in Eq.~\eqref{eq:b1}.
Following the steps above, and using the adversary matrix of 
Eq.~\eqref{eq:advmatrix},
we obtain
\begin{align}
\label{eq:inter2}
(1-\epsilon) n
   \le 4 \pi  \int_0^T dt \ \Delta(r(t))    \; .
\end{align}

We prove by contradiction 
that adiabatic approximations based on
local properties of the path cannot yield a schedule 
satisfying the local condition
$ \dot r(t) > c \Delta(r)/ \| \! \ket{\partial_r
\psi(r) } \! \| $, for some $c>0$ and all $r$.
If such condition is satisfied,
the inverse function 
$t(r)$ exists  and Eq.~\eqref{eq:inter2}
yields
\begin{align}
\label{eq:inter3}
(1-\epsilon) n
   < \frac{4 \pi} c  \int_0^1 dr \| \! \ket{\partial_r \psi(r)} \! \|= \frac{4 \pi L} c    \; .
\end{align}
Since $L =\pi n/2$ in this case, it is clear that $c<2 \pi^2/(1-\epsilon)$
or otherwise the above inequality is inconsistent.

{\bf Conclusions.} 
We proved that no general quantum process that
approximately prepares the eigenstate of a Hamiltonian
$H(1)$, by evolving with the path of Hamiltonians $\{ H(r) \}_{0\le r \le 1}$,
for any schedule $r(t)$, can achieve its goal in time less than
$\cO(L/\Delta)$.  The same bound on the time applies also
for more general evolutions with additional driving Hamiltonians.
Interestingly, some quantum processes for eigenpath traversal
almost achieve the bound under some assumptions~\cite{boixo_optimal_2009}.
We also gave a necessary local condition, valid even when the gaps of the Hamiltonians
and rates of change of the eigenstates are known along the path. In this case
we proved that if the schedule  satisfies $\dot r(t) > c \Delta(r) / \| \! \ket {\partial_r
\psi(r) }\! \| $, for a specific $c>0$ and all $r$,
the quantum process will not succeed in the state preparation. If only $\Delta \le \min_r \Delta(r)$ is known, our result suggests that no general monotonic schedule yields a cost better than $\cO(L/\Delta)$.

\begin{samepage}
\begin{acknowledgments}
We thank H. Barnum, R. Cleve, A. Childs, and E. Knill for discussions,
and the Perimeter Institute where part of this work was done.
SB thanks the National Science Foundation for support
under grant PHY-0803371 through the Institute for Quantum
Information at the California Institute of Technology.
RS thanks the Laboratory Directed Research and Development
Program at Los Alamos National Laboratory for support.
\end{acknowledgments}
\end{samepage}

%\bibliography{pathBound}

\end{document}